# Effect of Mn doping on ultrafast carrier dynamics in thin films of the topological insulator Bi$_2$Se$_3$

Yuri D. Glinka,[1,2]* Sercan Babakiray,[1] Mikel B. Holcomb,[1] and David Lederman[1]

[1]Department of Physics and Astronomy, West Virginia University, Morgantown, WV 26506-6315, USA
[2]Institute of Physics, National Academy of Sciences of Ukraine, Kiev 03028, Ukraine

Transient reflectivity (TR) measured at laser photon energy 1.51 eV from the indirectly intersurface coupled topological insulator Bi$_{2-x}$Mn$_x$Se$_3$ films (12 nm thick) revealed a strong dependence of the rise-time and initial decay-time constants on photoexcited carrier density and Mn content. In undoped samples ($x = 0$), these time constants are exclusively governed by electron-electron and electron-phonon scattering, respectively, whereas in films with $x = 0.013 – 0.27$ ultrafast carrier dynamics are completely controlled by photoexcited electron trapping by ionized Mn$^{2+}$ acceptors and their dimers. The shortest decay-time (~0.75 ps) measured for the film with $x = 0.27$ suggests a great potential of Mn-doped Bi$_2$Se$_3$ films for applications in high-speed optoelectronic devices. Using Raman spectroscopy exploiting similar laser photon energy (1.58 eV), we demonstrate that due to indirect intersurface coupling in the films, the photoexcited electron trapping in the bulk enhances the electron-phonon interaction strength in Dirac surface states.

## 1. Introduction

Ultrafast carrier dynamics in the topological insulator (TI) [1,2] Bi$_2$Se$_3$ is of great interest to scientific and technological communities due to potential applications of these materials in novel high-speed electronics. Owing to the unique nature of these three-dimensional (3D) materials that are insulating in the bulk (bandgap of Bi$_2$Se$_3$ ~ 0.3 eV) [3], but conducting at the surface due to two-dimensional (2D) Dirac surface states (SS), TIs offer an opportunity to design tunable electronic high-speed devices because of the different mechanisms of electron energy relaxation in 3D and 2D states associated with the electron-phonon (Fröhlich) interaction and deformation-potential/thermoelastic-scattering, respectively [4, 5].

Ultrafast carrier dynamics in TIs have been studied using time- and angle-resolved photoemission spectroscopy (TrARPES) [6-11], transient reflectivity (TR) at photon energy ~1.5 eV [4, 12-16], and ultrafast optical-pump terahertz-probe spectroscopy (probe photon energy ~1.0 meV) [17]. In the latter case, the ultrafast low-energy dynamics of free carriers residing in Dirac SS are found to take a few tens of ps, thus being much longer than the characteristic decay-time in graphene (1-2 ps) [18, 19]. Alternatively, the ultrafast carrier relaxation dynamics probed at photon energy ~1.5 eV includes fast electron-electron thermalization (~0.3-0.5 ps) and longitudinal-optical(LO)-phonon relaxation in the bulk (~1.5-3.5 ps) that leads to a metastable population of the conduction band edge [4, 12-16], which continuously feeds a non-equilibrium population of Dirac SS (~5-220 ps) [4,6,15] being likely mediated by the coherent acoustic phonon dynamics [20]. Afterwards the quasi-equilibrium carrier population in Dirac SS is reached within extremely long times of ≥10 ns [4, 8, 15, 20, 21]. In addition, recombination of carriers residing in the lower and higher energy Dirac cones [11] can strongly affect the carrier relaxation dynamics [15], giving rise to the broadband visible-range photoluminescence recently observed [22]. The ultrafast carrier dynamics can, in principle, also be affected by electron trapping when TIs are additionally doped with impurities. It would be of great interest, in this regard, to magnetic impurities (such as Mn) which can break down the time-reversal symmetry in Dirac SS and hence strongly affect the electron dynamics in TIs.

In this paper, we report on a TR study of ultrafast carrier relaxation dynamics probed at photon energy ~1.51 eV in the indirectly intersurface-coupled TI Bi$_{2-x}$Mn$_x$Se$_3$ films (12 nm thick). We show that Mn doping significantly alters the rise-time and initial decay-time constants of the TR signals, which in undoped samples are associated with electron-electron and electron-LO-phonon scattering, respectively [4, 12-16]. Consequently, ultrafast carrier relaxation in Mn-doped Bi$_2$Se$_3$ films is found to be governed by the electron trapping kinetics involving ionized Mn$^{2+}$ acceptors and their dimers. The shortest decay-time of ~0.75 ps observed for the film with $x = 0.27$ is similar to that of the characteristic value of massive Dirac fermion relaxation in bilayer graphene measured using TrARPES and exploiting optical pumping at similar photon energy (1.55 eV) [23], thus suggesting that Mn-doped Bi$_2$Se$_3$ thin films have a great potential for applications in high-speed optoelectronic devices.

## 2. Samples and experimental setup

Experiments were performed on 12 nm thick Bi$_{2-x}$Mn$_x$Se$_3$ epitaxial films with $x = 0, 0.013, 0.026, 0.078, 0.131, 0.182$, and $0.27$. The films were grown on a 0.5 mm thick Al$_2$O$_3$(0001) substrate by molecular beam epitaxy, with a 10 nm thick MgF$_2$ capping layer to protect against oxidation. The growth and Mn-doping processes were similar to that described previously [5, 24, 25]. The films were found to be naturally $n$-doped with an average free-electron density $n_e \sim 2.54 \times 10^{19}$ cm$^{-3}$, determined from Hall conductivity measurements at 2 K, which is typical for as-grown Bi$_2$Se$_3$ [26].

TR measurements were performed at room temperature in air using a Ti:Sapphire laser with a pulse duration of ~100 fs, a center photon energy of 1.51 eV (820 nm) and a repetition rate of 80 MHz. A wide range of laser powers (0.03-1.76 × 10$^9$ W/cm$^2$) was employed, where the pump was at normal incidence and the probe was at an incident angle of ~ 15°, focused through the same lens to a spot diameter of ~100 μm.

*Corresponding author: ydglinka@mail.wvu.edu

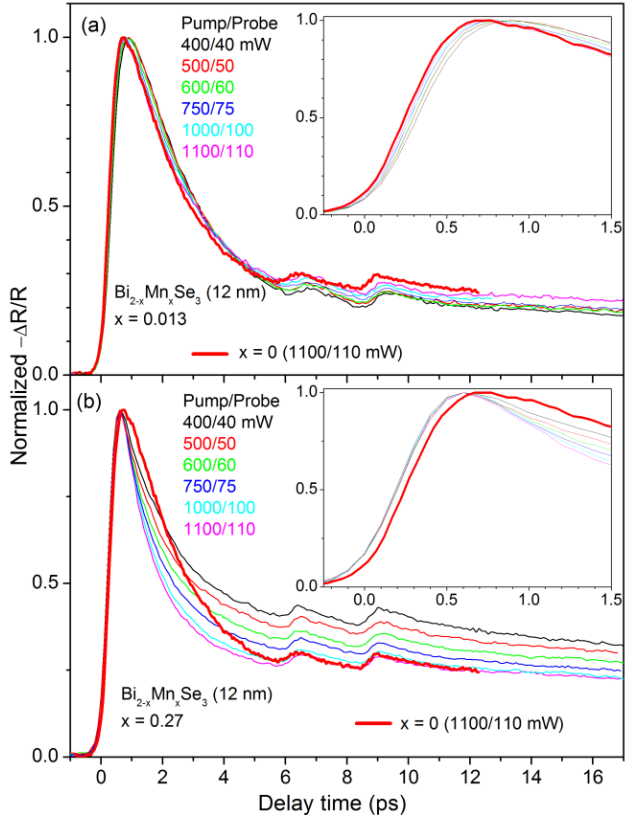

FIG. 1. Normalized TR traces for 12 nm thick $Bi_{2-x}Mn_xSe_3$ films with $x = 0.013$ (a) and $x = 0.27$ (b) measured with 0.05 ps steps and with various pump-probe average powers as indicated by the corresponding colors. Wider red curves correspond to the undoped sample ($x = 0$). Insets zoom in on the short delay-time part of the same TR traces.

The pump and probe beams were cross-polarized. No film damage was observed for the laser powers used in the measurements reported here. Raman measurements were performed also at room temperature in air using a Renishaw inVia Raman Spectrometer equipped with the 785 nm (photon energy 1.58 eV) solid-state laser of 30 mW output power [5, 26].

### 3. Experimental results and discussion

Figure 1 shows the typical TR traces measured with various pump-probe average powers for $Bi_{2-x}Mn_xSe_3$ films with $x = 0.013$ and $0.27$. The TR traces clearly demonstrate that the ultrafast carrier relaxation dynamics strongly depend on photoexcited carrier density and Mn content. Similar measurements were performed for undoped ($x = 0$) and other Mn-doped samples used in this study. We note that there are two weak stepwise features with onsets at ~5.9 and ~8.4 ps, which are due to the reflection of pump and probe beams from the internal surface of the sapphire substrate of thickness $d_{Sa} = 0.5$ mm, respectively. The amount of light transmitted through the sample is estimated as ~20 %. Due to the normal incidence of the pump, the first onset delay-time ($\tau_{onset}$) can be precisely obtained by taking into account the double-pass of incident pump light in the sapphire substrate of refractive index $n_r \sim$ 1.76, thus yielding $\tau_{onset} = 2d_{Sa}n_r/c = 5.9$ ps, where $c$ is the speed of light in vacuum.

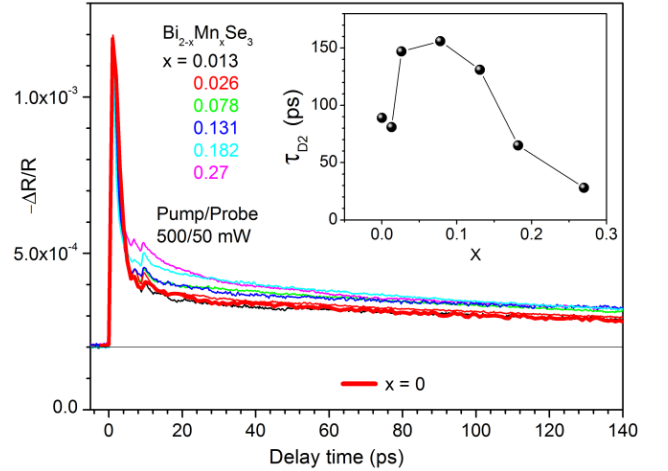

FIG. 2. TR traces for 12 nm thick $Bi_{2-x}Mn_xSe_3$ films measured with 1 ps steps and with various Mn content ($x$) as indicated by the corresponding colors. The wider red curve represents the TR trace of the undoped film ($x = 0$). Inset shows the Mn content dependence of $\tau_{D2}$ constant.

Figure 2 shows the longer delay-time TR traces that are negative in accordance with their absorption bleaching nature and are characterized by a rise and multiple decay behavior [4, 12-16]. The corresponding rise-time ($\tau_R$) and decay-time ($\tau_{Di}$) constants were estimated using a multiexponential fitting function [27],

$$-\Delta R/R = H(1-e^{-t/\tau_R})(a_1 e^{-t/\tau_{D1}} + a_2 e^{-t/\tau_{D2}} + a_3 e^{-t/\tau_{D3}}), \quad (1)$$

where $H$ is the Heaviside step function that accounts for the pump-probe cross-correlation time and $a_i$ are the partial amplitudes. For the undoped samples, $\tau_R$ is associated with electron-electron thermalization, whereas $\tau_{D1}$, $\tau_{D2}$, and $\tau_{D3}$ describe the intraband electron-LO-phonon relaxation in the bulk, the bulk interband electron-LO-phonon relaxation that feeds a population of Dirac SS, and a final localization of electrons in Dirac SS, respectively [4, 6, 8, 15, 16, 20, 21]. The doping of the films with Mn does not affect appreciably $\tau_{D3}$, whereas $\tau_{D2}$ reveals a non-monotonic behavior (figure 2, inset). Taking into account the bulk carrier depletion effect occurring with decreasing $Bi_2Se_3$ film thickness as a consequence of indirect intersurface coupling [15, 16], one can assume that an increase of $\tau_{D2}$ with increasing Mn content is due to the suppression of the interband electron-LO-phonon relaxation channel when the photoexcited electron density ($n$) in the bulk is reduced, being most likely due to trapping. Once Mn content exceeds $x \approx 0.1$, the opposite behavior emerges, indicating an increase of the bulk electron density. We note that because $n > n_e$ [4, 26], the effect of free-electron density on the photoexcited electron relaxation dynamics seems insignificant.

Similar non-monotonic behavior also appears for $\tau_R$ and $\tau_{D1}$. Figure 3 compares the variations of $\tau_R$ and $\tau_{D1}$ with increasing $n$ for undoped and Mn-doped $Bi_2Se_3$ films. In undoped films, $\tau_{D1}$ is associated with electron-LO-phonon interactions ($\tau_{e-LO} = 1.68$ ps), which remains unchanged with the pump power (with $n$), thus reflecting the fact that the rate of polar Fröhlich electron-LO-phonon interaction ($1/\tau_{e-LO}$) is independent of the carrier density [4]. One exception can occur for a degenerate electron gas in heavily doped semiconductors due to the energy band nonparabolicity, which mainly appears in low-temperature measurements [28]. In contrast, $\tau_R$ in undoped samples is associated with electron-electron interactions and hence slightly decreases with increasing $n$ (figure 3) since the electron-electron scattering rate ($1/\tau_R$) varies with carrier density as $n^{2/3}$ and as $n$ in 3D and 2D structures, respectively [29].

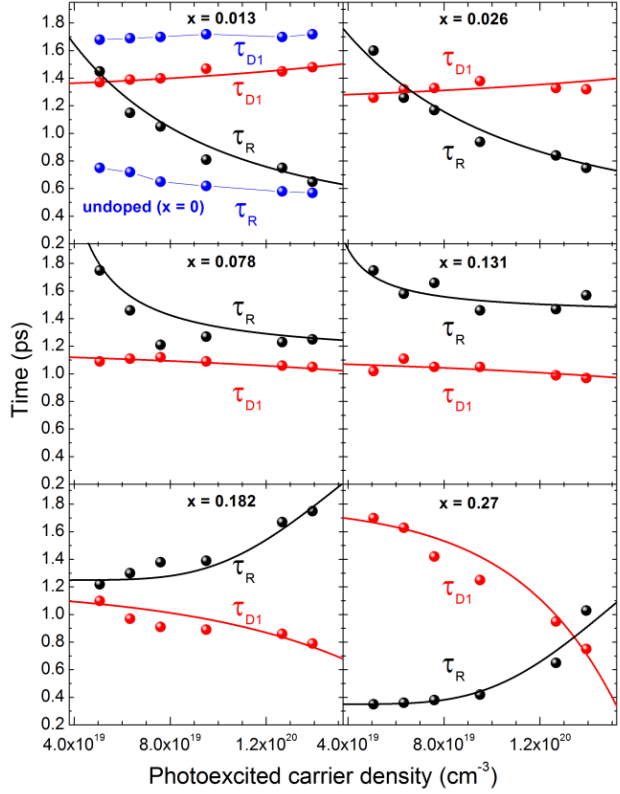
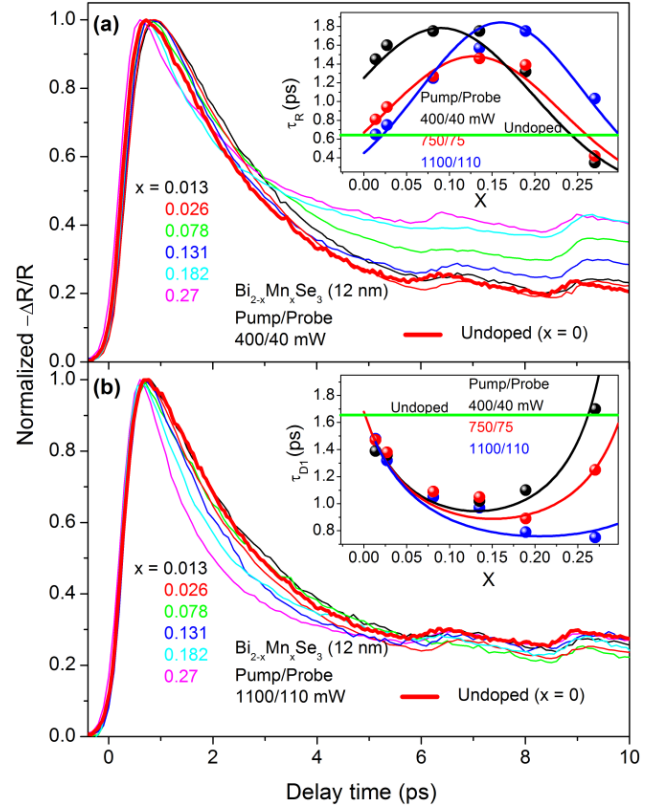

FIG. 3. The photoexcited carrier density dependences of $\tau_R$ and $\tau_{D1}$ for 12 nm thick $Bi_{2-x}Mn_xSe_3$ films with $x$ as indicated. The black and red color curves show the corresponding fits to the data using equations (8) and (10). All the fitting parameters are the same including $\tau_{e-LO} = 1.68$ ps, except for $A$, $\sigma_1$, and $\sigma_2$.

FIG. 4. Normalized TR traces measured with 0.1 ps steps and with 400/40 mW (a) and 1100/110 mW (b) pump-probe powers for 12 nm thick $Bi_{2-x}Mn_xSe_3$ films with $x$ as indicated by the corresponding colors. Insets show the Mn content dependences of $\tau_R$ and $\tau_{D1}$ for TR traces measured with three different pump/probe powers as indicated. The green horizontal lines indicate $\tau_R$ and $\tau_{D1}$ for the undoped film ($x = 0$). The corresponding color curves show the fits to the data using equations (8) and (10). All the fitting parameters are the same including $\tau_{e-LO} = 1.68$ ps, except for $A$, $\sigma_1$, and $\sigma_2$.

The $\tau_R$ and $\tau_{D1}$ trends with increasing $n$ and Mn content show a completely different behavior in Mn-doped $Bi_2Se_3$ films (figures 3 and 4) and therefore should be associated with electron trapping kinetics in a similar manner to those occurring in low-temperature grown GaAs, where the much shorter ultrafast carrier dynamics compared to high-temperature grown GaAs are mainly governed by the trapping of electrons by As cluster point defects [30]. Specifically, $\tau_R$ and $\tau_{D1}$ become comparable, giving rise to a quite symmetric shape of TR traces in the short delay-time range when the highest laser power is applied to the $Bi_{2-x}Mn_xSe_3$ film with $x = 0.27$ [figures 1(b) and 4(b)]. The corresponding shortest $\tau_{D1}$ ~0.75 ps is comparable to the characteristic value of massive Dirac fermion relaxation in bilayer graphene (0.65 ps) measured using TrARPES with optical pumping at similar photon energy (1.55 eV) [23], thus suggesting a new highly promising material for electronic and optoelectronic applications. In addition, the slopes of $\tau_R$ and $\tau_{D1}$ with increasing $n$ change sign when Mn content exceeds $x \approx 0.1$ (figure 3). Moreover, $\tau_R$ and $\tau_{D1}$ at fixed $n$ depend on $x$ also non-monotonically, revealing opposite trends with a minimum in $\tau_{D1}$ coinciding with a maximum in $\tau_R$, both shifting towards higher Mn content with increasing $n$ (figure 4, insets). This correlation between $\tau_R$ and $\tau_{D1}$ seems unique due to the joint dependence of the electron trapping and electron-electron scattering rates on the density of photoexcited electrons [29, 31].

The nature of traps is associated with the substitution of Mn atoms into $Bi^{3+}$ sites with 2+ valence [25, 31]. As shown in figure 5, Mn content dependences of Raman peak intensity, shift, and linewidth are monotonic for the entire range of Mn doping used, thus confirming that the substitution of Mn atoms in $Bi_2Se_3$ films dominates over possible intercalation between the quintuple layers [25, 32]. The red-shift of the bulk phonon modes ($E^2_g$ and $A^2_{1g}$) with increasing Mn content is due to phonon softening resulting from a weakening of interlayer van-der-Waals forces due to Bi substitution by the smaller ionic radius Mn atom [33, 34]. For the same reason, the bulk phonon modes significantly broaden [figure 5(a)], as it typically occurs in strain-induced or disordered systems [35, 36]. The broadening of the surface H-mode associating with the continuous hexagonally arranged Se atom network of Dirac SS [37] is asymmetric due to the Fano resonance [5, 26], which is maximized for films with $x \approx 0.13$ [figure 5(b), inset]. This behavior in general follows the non-monotonic dependence of the rise-time and initial decay-time constants with increasing Mn content (figures 2 and 4, insets), indicating that electron trapping in the bulk enhances the electron-phonon interaction strength in Dirac SS, similarly as with decreasing film thickness [5, 26]. It should be noted that the latter effect can also be associated with the time-reversal symmetry breaking induced by Mn impurities residing in close proximity to the film surface. The resulting elimination of the topological edge states will turn the system into a trivial insulator, where electron-phonon coupling is expected to be stronger due to the Fröhlich-type interaction.

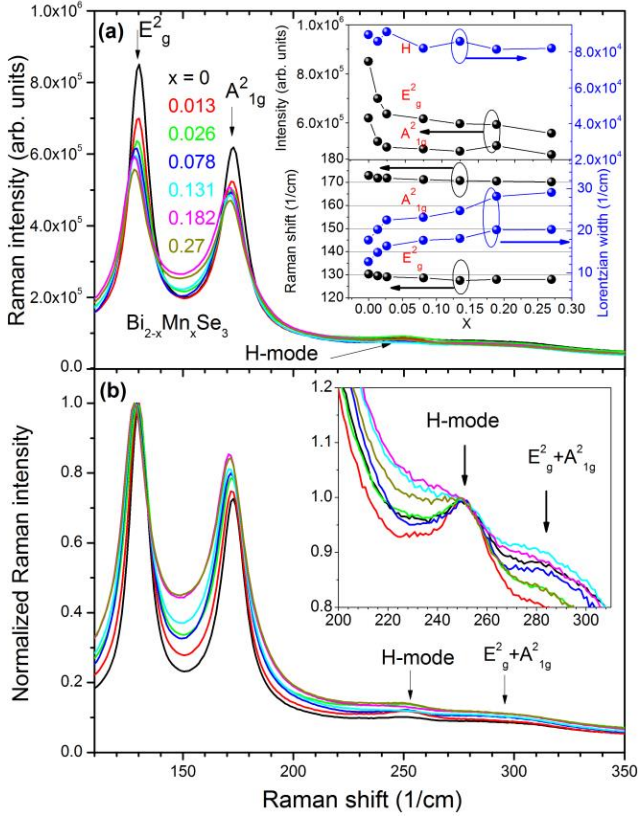

FIG. 5. (a) Raman spectra measured for 12 nm thick $Bi_{2-x}Mn_xSe_3$ films with $x$ as indicated by the corresponding colors. (b) The same spectra but normalized at the peak position of the bulk $E^2_g$ mode (~130 cm$^{-1}$). Inset in (a) shows the Mn content dependences of the phonon mode intensity, shift, and Lorentzian linewidth. Inset in (b) shows the same spectra but normalized at the peak position of the surface H-mode (~252 cm$^{-1}$).

Because $Mn^{2+}$ replaces a $Bi^{3+}$ ion, the substituted Mn atom provides one fewer electron or, equivalently, donates a hole, thus acting as an acceptor and showing a hole doping effect [38]. However, because the average vertical ionization energy of $Mn^{2+}$ (~1.6 eV) is comparable to the laser photon energy used in our experiments ($\hbar\omega_L$ = 1.51 eV) [39], all $Mn^{2+}$ acceptors are expected to be ionized to the $Mn^{3+}$ state ($Mn^{2+} \xrightarrow{\hbar\omega_L} Mn^{3+} + e$). Therefore, we consider $Mn^{3+}$ as electron traps in the photoexcited Mn-doped $Bi_2Se_3$ films, which after electron capture transform to $Mn^{2+}$ acceptor sites. This behavior is consistent with the generally accepted tendency regarding the compensation of donors in this family of materials by impurities with one fewer electron than Bi [40]. Furthermore, a weak coupling between two neighboring substituted $Mn^{2+}$ acceptors is expected to occur with increasing Mn content [41]. Two electron trapping by $Mn^{3+}$-$Mn^{3+}$ dimers immediately transform them to the strongest exchange interaction state of $Mn^{2+}$-$Mn^{3+}$ [42], donating one extra electron. Consequently, Mn dimerization reduces the electron trapping efficiency. We note that the substitutional-type aggregation of $Mn^{2+}$ acceptors should be distinguished from the Cu intercalation-type aggregation in the van der Waals gaps between the neighboring quintuple layers of Cu-doped $Bi_2Se_3$, where Cu atoms behave as donors [43].

The resulting evolution of the electron density is governed by the first-order trapping kinetics [44],

$$dn/dt = -\upsilon\sigma_1 n(n_t - n_t^*) + \upsilon\sigma_2 n(n_d - n_d^*), \qquad (2)$$

where $n$ and $n_t^*$ are the densities of photoexcited electrons and electrons trapped by ionized $Mn^{2+}$ acceptors of density $n_t$; while $n_d$ and $n_d^*$ denote the densities of dimers and the corresponding trapped electrons; $\upsilon$ is the electron thermal velocity and $\sigma_1$ and $\sigma_2$ are the trapping cross sections of the Mn monomers and dimers. The dimer density is expected to grow quadratically with $n_t$ [45, 46], so that

$$n_d = K n_t^2, \qquad (3)$$

where $K$ is a ratio of rate constants for dimers and monomers at a monomer-dimer self-associating equilibrium. Consequently, the ratio $K$ characterizes the efficiency of dimerization and has a dimension of cm$^3$. The solution of equation (2) is

$$n = n_0 e^{-t/\tau_t}, \qquad (4)$$

where $n_0$ is the initial density of photoexcited electrons that are subjected to trapping with a trapping rate

$$\tau_t^{-1} = \upsilon\sigma_1(n_t - n_t^*) - \upsilon\sigma_2(n_d - n_d^*). \qquad (5)$$

The density of photoexcited electrons trapped by ionized $Mn^{2+}$ acceptors and their dimers can be obtained using the first-order trapping equation [47],

$$dn_t^*/dn = \sigma_1 \delta_p(n_t - n_t^*) \; ; \; dn_d^*/dn = \sigma_2 \delta_p(n_d - n_d^*), \qquad (6)$$

where $\delta_p$ is the characteristic interaction length (light penetration depth ~12 nm) [4]. The solution of equation (6) is

$$n_t^* = n_t(1 - e^{-\sigma_1 \delta_p n_0}) \; ; \; n_d^* = n_d(1 - e^{-\sigma_2 \delta_p n_0}). \qquad (7)$$

Using equations (5) and (7) and taking into account that $\tau_{D1}^{-1} = \tau_{e-LO}^{-1} + \tau_t^{-1}$, $\tau_{D1}$ can be expressed as

$$\tau_{D1} = \tau_{e-LO}\left[1 + \tau_{e-LO}(\upsilon\sigma_1 n_t e^{-\sigma_1\delta_p n_0} - \upsilon\sigma_2 K n_t^2 e^{-\sigma_2\delta_p n_0})\right]^{-1}. \quad (8)$$

The slope of $\tau_{D1}$ with respect to both $n_0$ at fixed $n_t$ and $n_t$ at fixed $n_0$ changes sign when Mn dimers dominate the trapping kinetics. In the absence of dimerization ($K = 0$), $\tau_{D1}$ increases with increasing $n_0$ at fixed $n_t$ and decreases with increasing $n_t$ at fixed $n_0$, thus showing the common monomer-type electron trapping dynamics (figures 3 and 4).

Using equation (4), the electron-electron scattering rate for the 3D structure case [29] can be treated at fixed time $t_f$ as

$$\tau_R^{-1} = (1/A)n^{2/3} = (1/A)\left(n_0 e^{-t_f/\tau_t}\right)^{2/3}, \quad (9)$$

where $A$ is a proportionality coefficient. Accordingly, $(\tau_R)^{-1}$ is completely governed by electron trapping and Mn dimerization since both processes affect the electron density. Using equations (5), (7), and (9), the expression for $\tau_R$ finally reads as

$$\tau_R = A\left[n_0 e^{-\tau_f \upsilon(\sigma_1 n_t e^{-\sigma_1\delta_p n_0} - \sigma_2 K n_t^2 e^{-\sigma_2\delta_p n_0})}\right]^{-2/3}. \quad (10)$$

We note that $\tau_R$ changes the slope with respect to both $n_0$ at fixed $n_t$ and $n_t$ at fixed $n_0$ at the same conditions mentioned for $\tau_{D1}$, whereas the overall trend is opposite to that of $\tau_{D1}$. The fits of the experimental data shown in figures 3 and 4 using equations (8) and (10) reproduce all the variations of $\tau_{D1}$ and $\tau_R$ with $n_0$ and $n_t$, thus proving the validity of the model. We note that if the electron-electron scattering rate in Eq. (9) switches to the 2D structure case ($\tau_R^{-1} \sim n$) [29], the experimental data can also be fitted quite well, thus indicating that Mn doping effect on electron-electron thermalization dynamics is not sensitive enough to distinguish between 2D and 3D states in TIs. The Mn dimer cross section ($\sigma_2$) obtained from the fits is only ~3 times greater than that for Mn monomers ($\sigma_1 = \pi R^2$ ~0.03 nm$^2$, where $R$ ~0.1 nm is the Mn ionic radius), confirming that only one ionized Mn$^{2+}$ acceptor of the dimer traps the electron. Because $\sigma_2 > \sigma_1$, the Mn content extrema should shift toward the higher values with increasing $n_0$ due to the faster decrease of the negative term in the parentheses of equations (8) and (10), thus being in agreement with experimental observations shown in insets of figure 4.

## 4. Conclusion

In summary, we have demonstrated that Mn doping of 12 nm thick Bi$_2$Se$_3$ films strongly affects the rise-time and initial decay-time constants of photoexcited electron relaxation, which in undoped samples are exclusively governed by the electron-electron and electron-LO-phonon scattering, respectively. The effect is completely controlled by the kinetics of photoexcited electron trapping by ionized Mn$^{2+}$ acceptors and their dimers. The observed shortest decay-time of ~0.75 ps is comparable to the characteristic value of massive Dirac fermion relaxation in bilayer graphene, suggesting a great potential of Mn-doped Bi$_2$Se$_3$ films for applications in high-speed optoelectronic devices.

## Acknowledgments

We thank T. A. Johnson for helping with the growth of the MgF$_2$ capping layer in some of the original samples. This work was supported by a Research Challenge Grant from the West Virginia Higher Education Policy Commission (HEPC.dsr.12.29). Some of the work was performed using the West Virginia University Shared Research Facilities.